\newcommand{\nova}{NO$\nu$A}

\ifdefined\BW
  \newcommand{\sigstyle}{solid black}
  \newcommand{\ncstyle}{solid gray}
  \newcommand{\numustyle}{dashed black}
  \newcommand{\beamstyle}{dashed gray}
\else
  \newcommand{\sigstyle}{red}
  \newcommand{\ncstyle}{blue}
  \newcommand{\numustyle}{black}
  \newcommand{\beamstyle}{magenta}
\fi

\documentclass[preprint,5p,times,twocolumn]{elsarticle}

\usepackage{graphicx}
\graphicspath{{figures/}}

\usepackage{appendix}

\journal{Nuclear Instruments and Methods A}

\begin{document}

\begin{frontmatter}

\title{Library Event Matching event classification algorithm for electron neutrino interactions in the \nova{} detectors}

\author{C. Backhouse}
\author{R. B. Patterson}
\address{Lauritsen Laboratory, California Institute of Technology, Pasadena, California 91125, USA}

\author{}

\address{}

\begin{abstract}
We describe the Library Event Matching classification algorithm implemented for use in the \nova{} $\nu_\mu\,\mathord{\to}\,\nu_e$ oscillation measurement.  Library Event Matching, developed in a different form by the earlier MINOS experiment, is a powerful approach in which input trial events are compared to a large library of simulated events to find those that best match the input event.  A key feature of the algorithm is that the comparisons are based on all the information available in the event, as opposed to higher-level derived quantities.  The final event classifier is formed by examining the details of the best-matched library events.  We discuss the concept, definition, optimization, and broader applications of the algorithm as implemented here.  Library Event Matching is well-suited to the monolithic, segmented detectors of \nova{} and thus provides a powerful technique for event discrimination.
\end{abstract}


\end{frontmatter}

\section{Introduction}

Classifying images into a small number of categories is a common task in scientific and industrial fields.  In particle physics, this task usually involves interpreting particle detector data to determine the type of particles, interactions, or decays present.  Given the sheer volume of information that can be collected, the data is often first reduced to a set of derived quantities by running algorithms that pull out key features:\ clusters, tracks, showers, jets, etc.  While this form of lossy compression is acceptable in some applications, it is worth exploring whether a classification scheme that uses all of the available information is feasible, even in cases where the data volume is high.

In this article we describe such a classification scheme developed to categorize neutrino scattering events recorded in the \nova{} detectors.  In the Library Event Matching (LEM) algorithm, a trial event of unknown type is compared to a large number of known ``library'' events to find those events that are most similar to the trial event.  The properties of those best-matched library events reveal the likely nature of the trial event.  A distinguishing feature of LEM is that the comparisons are made using the energy depositions directly, to avoid any information loss from calculating higher-level variables.  This fundamental philosophy of LEM was developed within the MINOS collaboration for its own neutrino event categorization needs~\cite{minosnue1,minosnue2,pedro,ruth}.  The LEM version described in this article has substantial differences from its predecessor, many of which are motivated by the higher spatial resolution of the \nova{} detectors.

While we use \nova{} as our case study, the approach discussed is generalizable and could be usefully applied to any highly segmented detector, from hadron calorimeters determining jet multiplicity to cubic kilometer arrays collecting neutrinos from astrophysical sources.  As with many machine learning algorithms, LEM requires a large number of known examples from each classification category.  In particle physics applications, these would typically come either from an advanced Monte Carlo simulation or from calibration sources.

\begin{figure}
  \includegraphics[viewport=-100 0 1850 1280,width=0.95\linewidth]{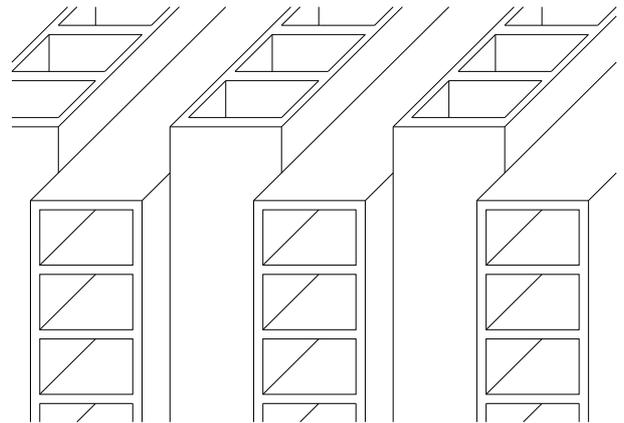}
  \caption{A sketch of the structure of the \nova{} detectors.  4~cm~$\times$~6~cm cells run the length of each 16~m~$\times$~16~m plane. The alternating vertical and horizontal orientations can be seen.  They are filled with liquid scintillator and each contains a looped wavelength-shifting fiber (not shown), as described in the text.  This cut-away sketch is diagrammatic only.  The real cells have rounded corners and the ends of the cells are capped for instrumentation and oil containment purposes.  The neutrino beam is incident from the left.}
  \label{fig:det}
\end{figure}

\begin{figure*}
  \begin{minipage}{.5\linewidth}
    \ifdefined\BW
      \includegraphics[trim=10 20 30 35, clip, width=\linewidth]{evd_nue_bw}
    \else
      \includegraphics[trim=10 20 30 35, clip, width=\linewidth]{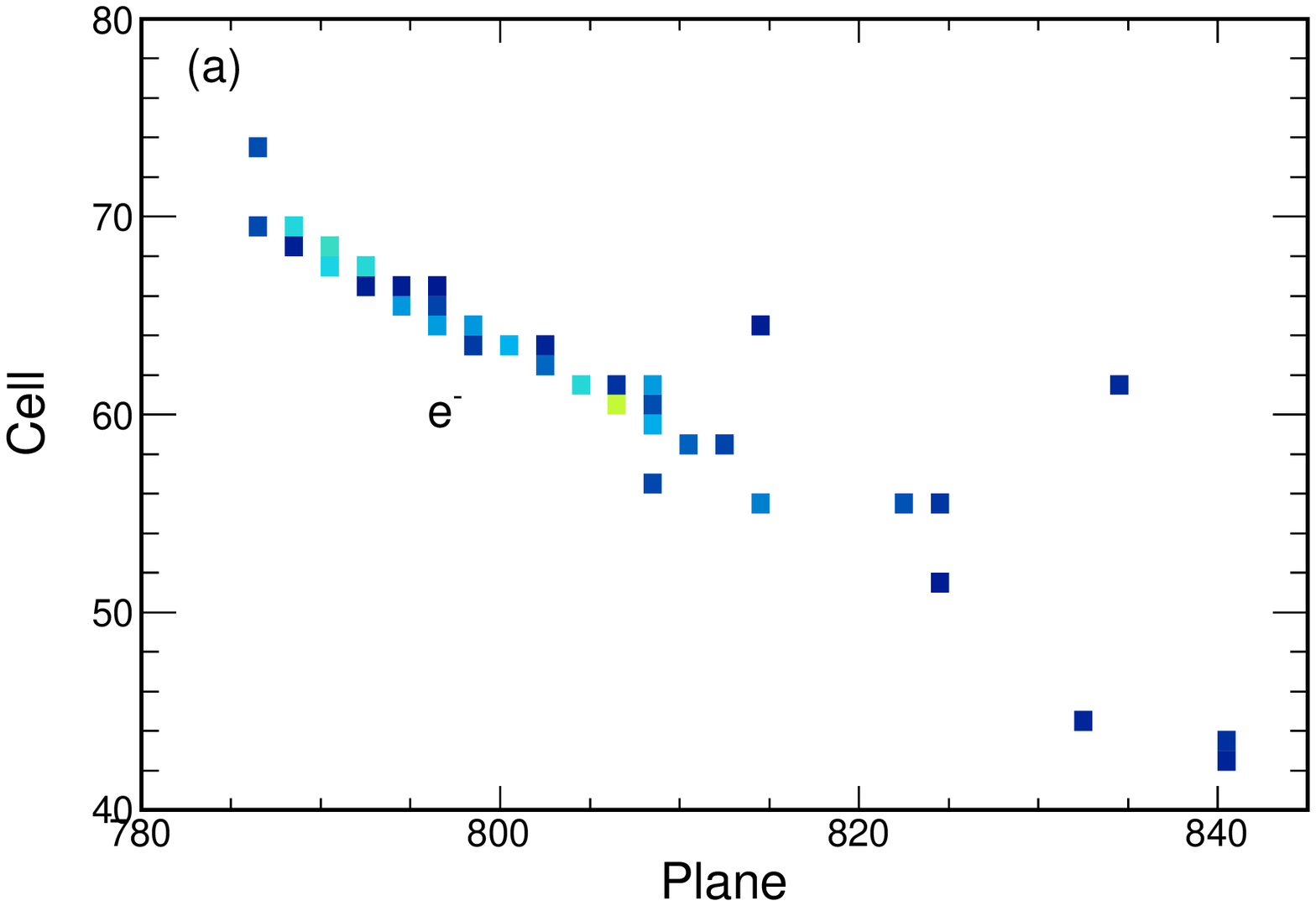}
    \fi
  \end{minipage}
  \begin{minipage}{.5\linewidth}
    \ifdefined\BW
      \includegraphics[trim=10 20 30 35, clip, width=\linewidth]{evd_nc_bw}
    \else
      \includegraphics[trim=10 20 30 35, clip, width=\linewidth]{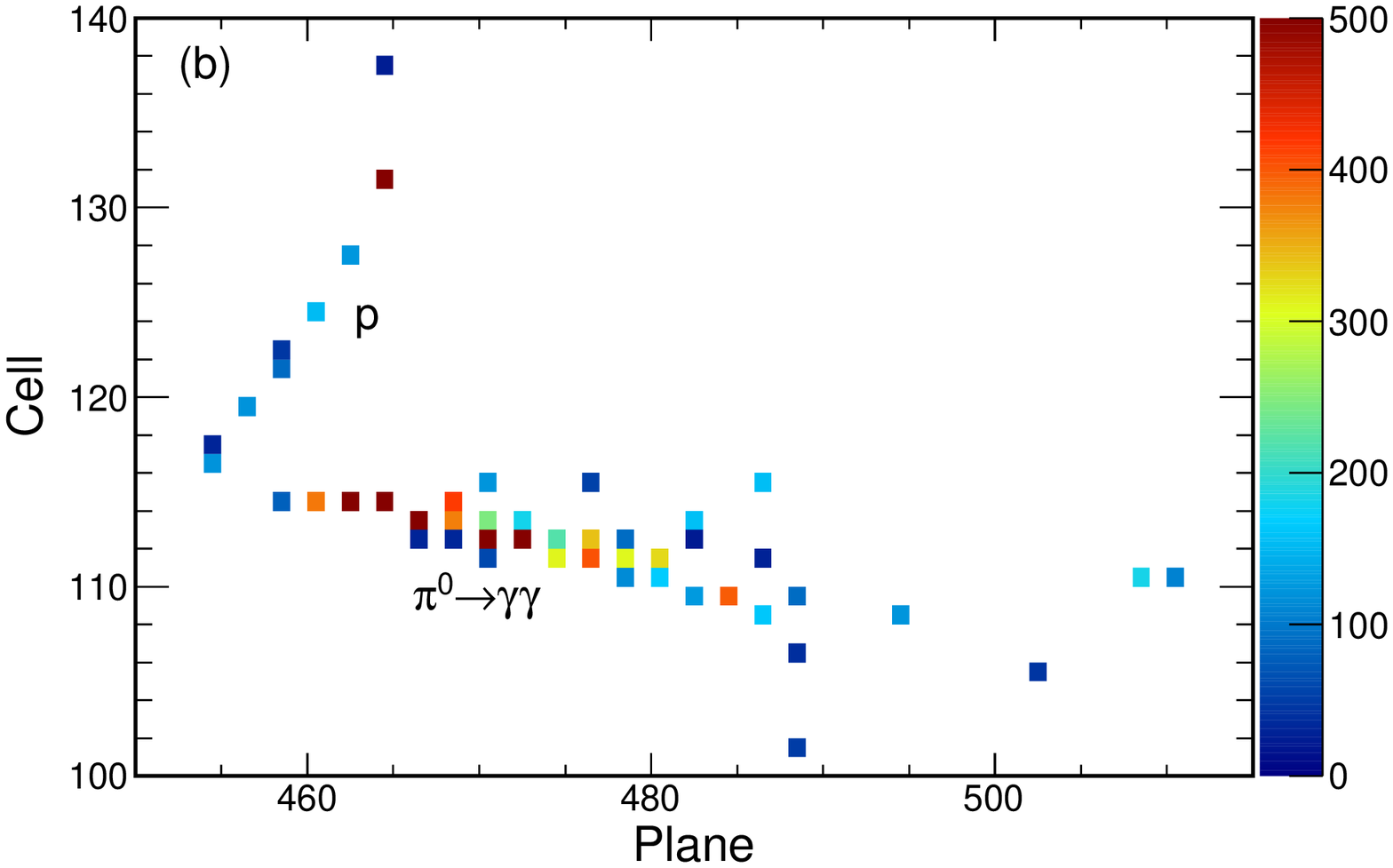}
    \fi
  \end{minipage}\ \\\ \\\ \\
  \begin{minipage}{.5\linewidth}
    \ifdefined\BW
      \includegraphics[trim=10 20 30 35, clip, width=\linewidth]{evd_cc_bw}
    \else
      \includegraphics[trim=10 20 30 35, clip, width=\linewidth]{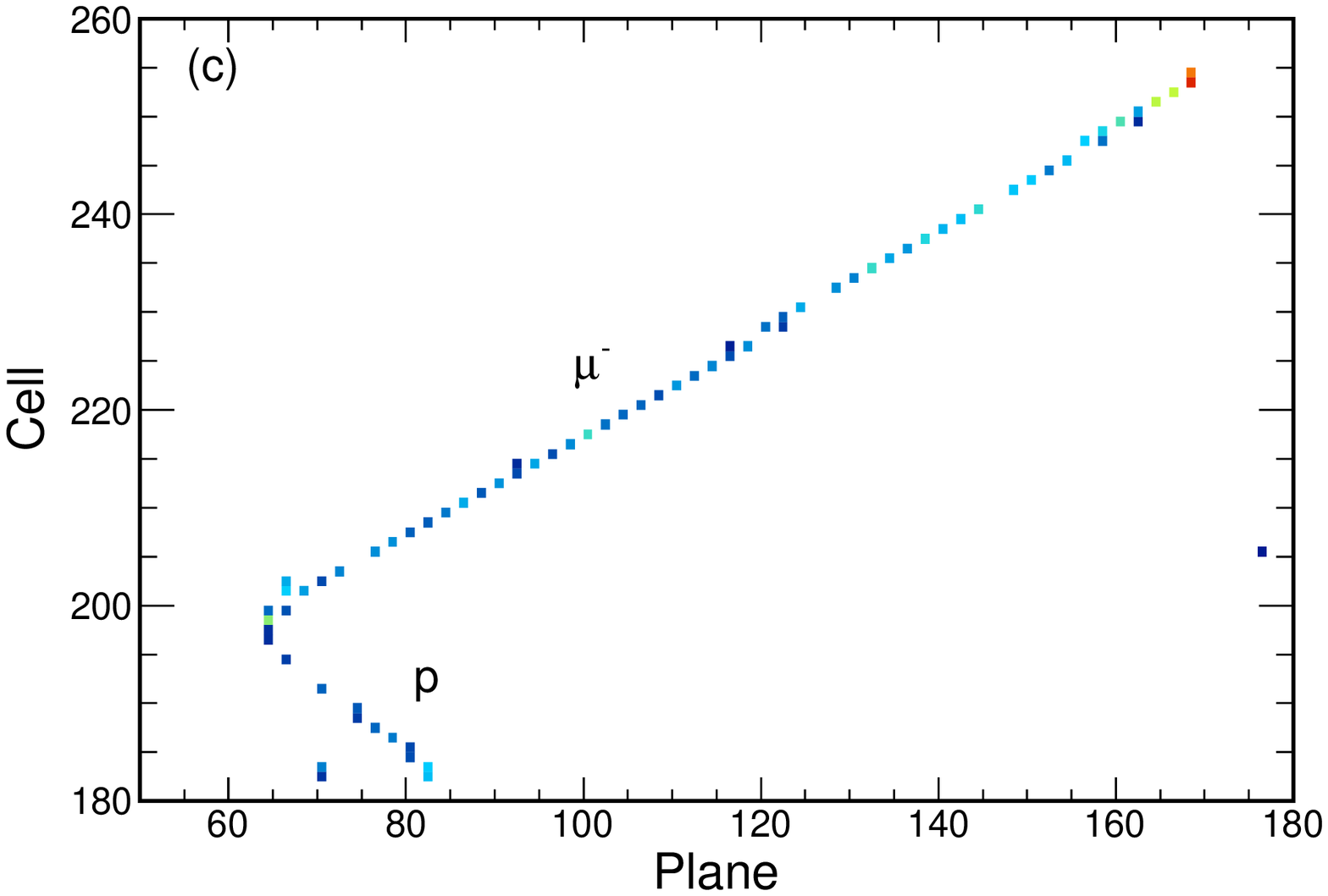}
    \fi
  \end{minipage}
  \begin{minipage}{.5\linewidth}
    \caption{Example simulated events in the \nova{} detectors.  Only one of the two views is shown in each case.  Each box represents one cell and is positioned according to its plane number (horizontal axis) and cell number (vertical axis).  The color scale indicates the charge deposited in photoelectrons, and is common to all three panels.  (a) A $\nu_e$ CC event, with the electron-induced electromagnetic shower clearly visible. (b) A neutral current event with a $\pi^0$.  The upper track is due to a proton.  This event shows that the two showers from $\pi^0\,\mathord{\to}\,\gamma\gamma$ are not always distinct.  (c) A $\nu_\mu$ CC event, with the usual tell-tale long, straight muon track. Note that the axis ranges are approximately doubled for this panel relative to the first two.}
    \label{fig:evds}
  \end{minipage}
\end{figure*}

\section{The \nova{} experiment}

The \nova{} (NuMI Off-axis $\nu_e$ Appearance) experiment studies the phenomenon of neutrino flavor oscillation~\cite{nova}.  Neutrinos produced by the NuMI beamline at Fermilab~\cite{numi} are observed by a Near Detector on the Fermilab site and by a Far Detector of identical construction located 810~km downstream in Ash River, Minnesota.  For the purposes of this article, the neutrino oscillation mode of interest is $\nu_\mu\,\mathord{\to}\,\nu_e$, and the goal of the classification algorithm is to obtain a sample of electron neutrino interactions in the Far Detector with the highest possible efficiency and purity.

The \nova{} detectors are constructed from long PVC cells filled with scintillator-doped mineral oil.  Each of the Far Detector's 344,064 cells is 16~m long with rectangular cross section 4~cm~$\times$~6~cm.  A loop of wavelength-shifting fiber runs the length of each cell, with both ends of the fiber terminating at one pixel of a 32-pixel APD array.  The body of the 14-kiloton detector consists of 896 layers, or ``planes'', each with 384 cells.  Each plane is 16~m~$\times$~16~m square, and the depth of the detector along the beam direction is 60~m.  Alternate planes are aligned vertically and horizontally so that three-dimensional information can be obtained through combination of the two ``views''.  The detector has unprecedented granularity for its size, with one radiation length (38~cm) extending over many cells, to give a detailed view of neutrino-induced electromagnetic showers.  Figure~\ref{fig:det} shows a cut-away diagram of the detector's construction.

The signal for the $\nu_\mu\,\mathord{\to}\,\nu_e$ oscillation analysis in \nova{} is $\nu_e$ charged-current (CC) scattering, which yields a high-energy electron in the final state that allows one to tag the incident neutrino's flavor.  In the 1 to 3~GeV energy range of \nova{}, this electron will be accompanied, with similar probabilities, by a proton (quasi-elastic scattering), a nucleon plus a pion (resonant scattering), or a richer hadronic shower (deep inelastic scattering).  While nuclear effects blur these crisp definitions, these three scattering types are useful for conveying the variety of shapes that signal events in \nova{} can take.  The $\mathord{\sim}1\ \mathrm{GeV}$ electron in the final state produces an electromagnetic shower in the detector that has a width of a few cells and runs longitudinally an average distance of 2.5~m (40 planes).  Figure~\ref{fig:evds}a shows a simulated $\nu_e$ CC interaction in the \nova{} Far Detector.

The primary mis-identification background comes from neutral-current (NC) interactions, particularly those where the recoil hadronic system contains a $\pi^0$.  The $\pi^0$ decays quickly to two photons, each of which induces an electromagnetic shower that is essentially indistinguishable from an electron-induced shower.  NC $\pi^0$ events, taken as a whole, look sufficiently different from signal $\nu_e$ CC events that we can reject them well, but the differences are sometimes obscured:
\begin{itemize}
\item The presence of two electromagnetic showers, rather than one, can reveal a $\pi^0$ in the final state.  However, if one of the showers has low energy or overlaps the other in the detector, it can be missed.
\item Photon-induced showers are separated from the neutrino interaction point due to the distance traveled by the photon prior to its conversion.  This gap is a tell-tale sign of a photon, but in some cases the gap will be too small to resolve.  The conversion length in \nova{} is 50~cm.
\item Photon-induced showers begin with two particles (an electron/positron pair) rather than one, but these cases can end up indistinguishable given the energy resolution of the detector.
\item The energy lost to the outgoing neutrino in NC scattering leads to reconstructed energies lower than those of signal events.  However, interactions from a sufficiently high-energy neutrino or with a large energy transfer can fall in the signal region of 1 to 3~GeV reconstructed energy.
\end{itemize}
Figure~\ref{fig:evds}b shows a simulated NC event with a $\pi^0$.

Additional background comes from $\nu_\mu$ CC scattering, which produces a muon in the final state.  The muon leaves a long track of activity in the detector with a characteristic energy deposition per unit pathlength.  These are readily removed from the sample due to the clear muon track except in cases where the muon is low in energy or is lost amongst other activity.  In these cases, the background is similar to NC interactions, with neutral pions playing the same role.  Figure~\ref{fig:evds}c shows a $\nu_\mu$ CC example.

\begin{figure}
  \ifdefined\BW
    \includegraphics[viewport=0 0 523 353,clip=true,width=\linewidth]{espec_bw}
  \else
    \includegraphics[viewport=0 0 523 353,clip=true,width=\linewidth]{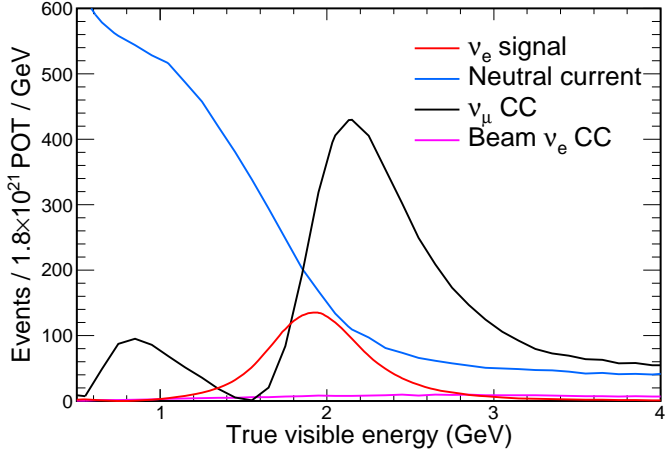}
  \fi

  \caption{Signal and background distributions of visible energy expected in the Far Detector sample. The effect of neutrino oscillations is included.  Visible energy is defined as the incident neutrino energy except in the case of neutral current events where the outgoing neutrino energy is subtracted.  The $\nu_e$ signal to be identified by LEM is shown in \sigstyle{}. The
    neutral current, $\nu_\mu$ charged current, and intrinsic beam $\nu_e$
    charged current components are \ncstyle{}, \numustyle{}, and \beamstyle{} respectively.
  }
  \label{fig:espec}
\end{figure}

The NuMI beam also includes a 2\% contamination of $\nu_e$.  These $\nu_e$ interact identically to the $\nu_e$ from oscillations and thus constitute a background to the $\nu_\mu\,\mathord{\to}\,\nu_e$ oscillation measurement.  However, their rate is low and their energies are somewhat higher.  Figure~\ref{fig:espec} illustrates the energy differences among all the event classes before any selection cuts have been applied.

Since the $\nu_e$ CC signal falls within a known energy range, we can safely remove lower and higher energy events up front.  For all figures and tables that follow, we require events to have reconstructed visible energies between 0.5~GeV and 4~GeV.

\section{Library Event Matching concept}

At the heart of the LEM algorithm is the comparison of each unknown trial event to a large number of known library events, with the comparisons based on low-level information collected by the detector.  For \nova{}, this means using the calibrated energy depositions in all the detector cells directly rather than forming higher-level objects such as showers and tracks from those.

Once the very best matches are found (here, the best 0.0001\% of all library events), their known properties are used to estimate the properties of the trial event.  In the simplest version of LEM, the fraction of the best matches that are signal events can be used as the discriminant.  \ref{sec:knnsvm} discusses the relationship between LEM and other machine learning techniques.

\subsection{The matching metric: motivation}
When comparing two events, a metric is needed to quantify how similar they are.  It is instructive to look at the MINOS case briefly, as the situation there is somewhat simpler~\cite{minosnue1,minosnue2,pedro,ruth}.

The MINOS detector has a segmented structure analogous to that of the \nova{} detector, but the effective spatial resolution for events of interest is significantly lower.  A $\nu_e$ CC signal event in MINOS involves only a couple dozen active ``strips'' (the analogue of \nova{}'s cells), and these active strips are clustered in a relatively compact pattern.  Thus, two events with the same underlying particle kinematics have a good chance of having identical (or near-identical) arrangements of active strips.  The readout electronics report the number of photoelectrons detected in each active strip.  Since this charge measurement suffers from shot noise (typical charge:\ $\sim$8~photoelectrons), strips with identical energy depositions may report different charges.  The level of difference is governed by Poisson statistics.

These details guided the form of matching metric used by MINOS, which can be thought of as the likelihood $\mathcal{L}$ that the two events' recorded charges represent the same underlying energy depositions:
\begin{equation}
  \log\mathcal{L} = \sum_i^{\rm strips} \log \left[\int P(a_i|\lambda)P(b_i|\lambda) d\lambda\right]\ ,
\end{equation}
where $a_i$ is number of photoelectrons registered by the $i^{\mathrm{th}}$ strip of event A, $b_i$ is the same for event B, $P(n|\lambda)$ is the Poisson probability of observing $n$ given mean $\lambda$, and the sum runs over all strips active in at least one of the events.  A higher $\log{\mathcal{L}}$ for a pair of events means a better match.  Before $\mathcal{L}$ is calculated, the events, which in general occur in different parts of the detector, are spatially aligned by shifting them so that their charge-weighted mean strip positions, rounded to the nearest strip, overlap.

In the MINOS metric $\mathcal{L}$, displaced energy depositions in the two events do not get their charges directly compared.  To obtain good matches for a trial event, the library must be large enough to span minor variations in active strip positions for nominally equivalent events.  This is possible in MINOS given the limited spatial resolution of the detectors for $\nu_e$ CC events.  That is, the library can be expected to give reasonable coverage of all possibilities.  Requiring exact {\em charge} agreement across the $\sim$20 active strips, though, would be combinatorically overwhelming.  The Poisson factors take care of this, with acceptably different charges able to contribute appropriately to the match score.

The \nova{} detectors are significantly more finely-grained than those of MINOS.  This makes event discrimination easier in principle since more details are visible, but it makes the above matching metric impractical.  It is much less likely that ``equivalent'' activity in the trial and library events will fall on the same cells.  What is needed is a matching metric that rewards activity in nearby cells without requiring them to lie directly on top of one another.  A library event identical to the trial event should still be a perfect match, but events with similar charges offset by a cell or so should still score well.

\begin{figure*}
  \begin{center}
    \ifdefined\BW
      \includegraphics[viewport=0 14 568 365,clip=false,width=0.93\linewidth]{match_images2_bw}
    \else
      \includegraphics[viewport=0 14 568 365,clip=false,width=0.93\linewidth]{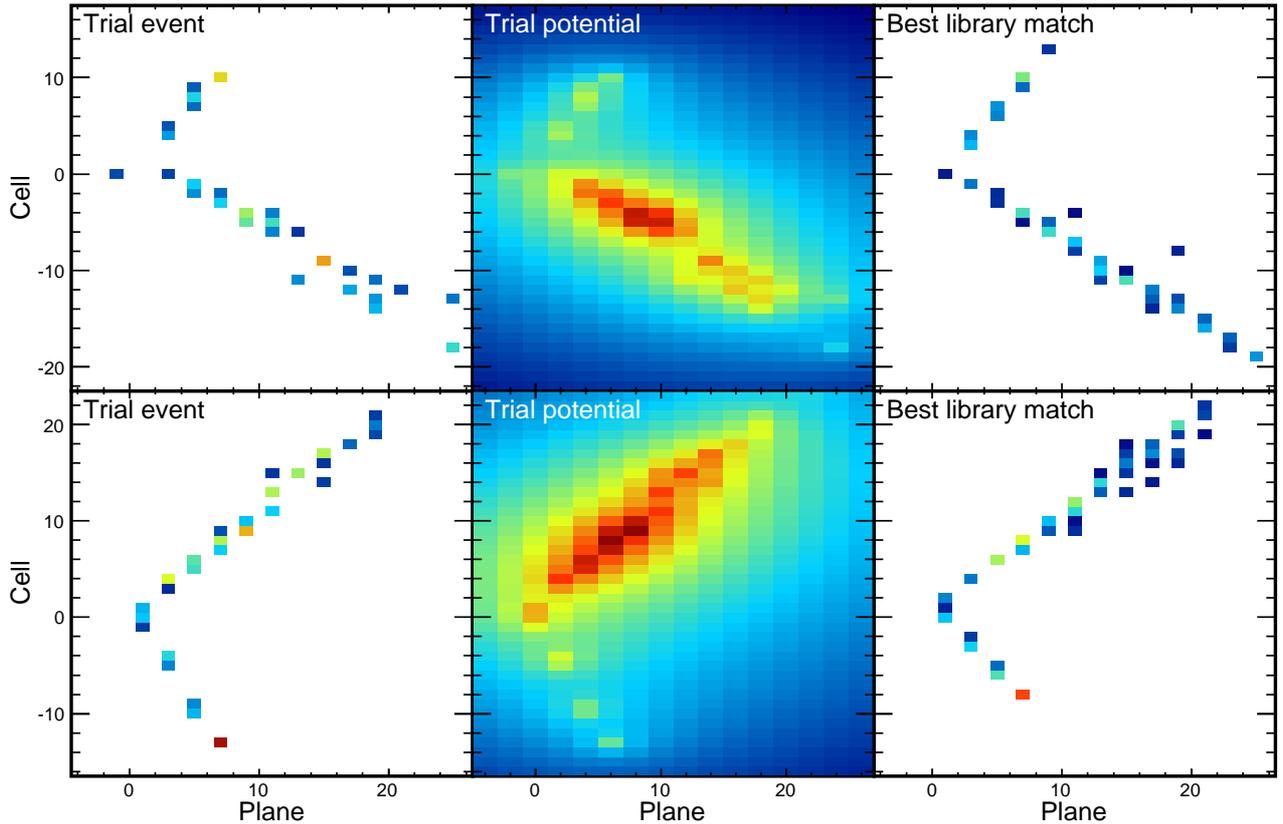}
    \fi
  \end{center}
  \caption{Example of LEM matching. On the left is a trial $\nu_e$ CC event, on
  the right the best match found. The central panels shows the potential $U$ in
  which the library events are placed in order to calculate the match
  energy. The upper panels show one view, and the lower panels show the other.}\label{fig:matchexample}
\end{figure*}

The metric we use draws its motivation from electrostatics. Two Coulomb charge distributions of similar shape, but with opposite signs, will have a low electrostatic potential energy when overlaid and examined together, as the attraction between the opposite signed charges counters the internal repulsion of the like-signed charges. Two overlaid charge distributions with dissimilar shape suffer the internal repulsion but lack the benefit of mutual attraction, leading to a large potential energy.  Given the electrostatic analogue to what follows, we use ``energy'' to refer to the LEM match score for the remainder of the article unless otherwise stated.  Lower energies correspond to better matches.

The match energy is defined as
\begin{equation}\label{eqn:potential}
E = E_A + E_B + E_{AB}\ ,
\end{equation}
where $E_A$ is the self-energy (repulsion) of event A's charges, $E_B$ is the self-energy of event B's charges, and $E_{AB}$ is the (negative) energy due to the the A/B attraction.  The charges are taken to be the recorded energy depositions in the \nova{} cells.  Treating the electrostatic analogue as exact for a moment, the self-energy terms are given by
\begin{equation}
E_A = \frac{1}{2}\sum_{ij}^{\rm cells}\frac{a_ia_j}{r_{ij}}\ ,
\ \ \ \ \ \ E_B = \frac{1}{2}\sum_{ij}^{\rm cells}\frac{b_ib_j}{r_{ij}}\ ,
\end{equation}
with $a_i$ ($b_i$) the recorded deposition in the $i^{\mathrm{th}}$ cell of event A (event B) and with $r_{ij}$ the distance between cells $i$ and $j$.  The $r_{ij}=0$ case is handled again with an electrostatic analogue by distributing all charges uniformly across their individual cells.  (See \ref{sec:zeror}.)

The interaction term is given by
\begin{equation}\label{eqn:interaction}
E_{AB} = -\sum_{ij}^{\rm cells}\frac{a_ib_j}{r_{ij}}\ .
\end{equation}
Before evaluating this sum, the events are globally aligned with one another according to a separately reconstructed interaction vertex.\footnote{Alignment by charge-weighted mean cell position was also studied and gives similar classification performance.}

A perfect match, in which events A and B have identical depositions in identical cell positions, would yield $E\mathord{=}0$.  A poorly matched pair with charges far away from one another will have large energy $E\approx E_A + E_B$.

Eq.~(\ref{eqn:interaction}) can be recast in terms of one set of charges embedded in the field of the other:
\begin{eqnarray}
  E_{AB} &=& -\sum_i^{\rm cells} a_iV_i \label{eqn:potential_voltage}\\
  {\rm where\ } V_{i} &=& \sum_j^{\rm cells} {b_{ij}\over r_{ij}} \label{eqn:voltage}\ .
\end{eqnarray}
The advantage of this formulation is that $V$ can be precalculated for each trial event, along with the self-energies of the trial and library events.  When matching against a large number of library events using (\ref{eqn:potential_voltage}), the complexity is linear in the number of
charges rather than requiring a double sum over both trial and library charges.

\subsection{The matching metric in \nova{}}

While the \nova{} matching metric is inspired by electrostatics, there is no reason to expect that the precise form above will yield the best sensitivity.  We incorporate the following generalizations.
\begin{itemize}
\item Above, $r_{ij}$ is calculated as the Euclidean distance in terms of the number of planes $\Delta p_{ij}$ and number of cells $\Delta c_{ij}$.  However, \nova{} events are boosted forward and cover many planes longitudinally but relatively few cells transversely, so we assign different relative importance to separations in the two directions.
\item The $r^{-1}$ falloff with distance is generalized to $r^{-\alpha}$.
\item The importance of larger charges relative to smaller ones is adjusted by raising all charges to a power $\beta$.
\end{itemize}
The resulting form of the matching metric still follows Eq.~(\ref{eqn:potential}), but the self-energy and interaction terms are now given by
\begin{eqnarray}
  E_A &=& \frac{1}{2}\sum_{ij}^{\rm cells}a_i^{\beta}T_{ij}a_j^{\beta}\ ,
\ \ \ \ \ E_B \ \ \ = \ \ \ \frac{1}{2}\sum_{ij}^{\rm cells}b_i^{\beta}T_{ij}b_j^{\beta}\\
  E_{AB} &=& -\sum_{i}^{\rm cells}a_i^{\beta}U_{i}
\end{eqnarray}
with the transfer matrix $T_{ij}$ and field $U_{i}$ given by
\begin{eqnarray}
  T_{ij} &=& \left(\frac{\Delta p_{ij}^2}{\sigma_p^2} + \frac{\Delta c_{ij}^2}{\sigma_c^2}\right)^{-\alpha/2} \label{eqn:transfer}\\
  U_i &=& \sum_{j}^{\rm cells}T_{ij}b_j^{\beta}\ .
\end{eqnarray}
The electrostatics version is recovered by setting
\begin{equation}
  \sigma_p = \sigma_c = \alpha = \beta = 1\ .
\end{equation}
We ran toy experiments with different values of these parameters and calculated a figure-of-merit for each to optimize performance.  The parameters chosen were:
\begin{eqnarray}
  \sigma_p &=&  0.286\\
  \sigma_c &=&  0.095\\
  \alpha   &=&  0.25\\
  \beta    &=&  0.5\ .
\end{eqnarray}
The first two parameters validate the intuition that transverse differences should be considered more significant than longitudinal ones.  The third parameter specifies a $1/\sqrt[4] r$ falloff with distance, slower than the electrostatic analogue.  For $\beta$, note that the simple presence or absence of activity in a cell conveys information regardless of its charge.  Having $0\mathord{<}\beta\mathord{<}1$ moves the metric towards this binary ``on/off'' interpretation and away from a charge-proportional weighting.

\section{The library}
The library consists of 77M simulated neutrino events, of which 18M are signal $\nu_e$ CC events, 29M are background $\nu_\mu$ CC and NC events, and 30M are $\pi^0$-enriched NC background events.  Each trial event that LEM classifies is compared to these 77M events to find the 1,000 library events that are most similar to it, as quantified by the metric above.\footnote{This statement is modified in Sec.~\ref{sec:heads} when we discuss speed optimizations.}  Figure~\ref{fig:matchexample} shows an example trial event along with its event potential $U$ and its best-matched library event.

The library events are generated ahead of time using the full \nova{} Monte Carlo simulation chain including realistic neutrino flux, cross sections, and detector components.  The flux is calculated using a FLUKA/FLUGG implementation of the beamline elements~\cite{fluka}, the neutrino interactions are simulated by GENIE~\cite{genie}, and particle propagation through the detector geometry is handled by GEANT4~\cite{geant}.  Simulated energy depositions in the liquid scintillator are converted into expected signals by \nova{} electronics and data acquisition simulation code.  The registered signals are corrected for light attenuation in the cells' fibers using standard \nova{} calibration procedures.

NC events containing neutral pions are the dominant mis-identification background owing to the electromagnetic showers from $\pi^0\rightarrow\gamma\gamma$.  Thus, we supplement the base background library sample with a $\pi^0$-enriched library sample.  To build this enriched sample, we apply a cut that selects out only those neutral current events with a $\pi^0$ present in the final state as reported by GENIE.

The library events are generated according to the expected $\nu_\mu$ flux (for background) or a 100\% $\nu_\mu\,\mathord{\to}\,\nu_e$ transmutation (for signal), without regard to any actual probabilities for neutrino flavor change.  Oscillations are introduced into the library later by event weighting.  This is discussed in Sec.~\ref{sec:mva} below.  \ref{sec:oscweights} describes the oscillation probabilities used.

While increasing the library size beyond the 77M events would provide incremental improvement in classification performance, we observe that these gains enter logarithmically with the number of library events once the library is sufficiently large.  In an earlier version of the algorithm, we found that doubling the library size provided only 1\% gain in physics sensitivity.  In light of the computational requirements discussed in Section~\ref{sec:comp}, additional library events are not worthwhile for our application.

\subsection{Event flipping}

To good approximation, flipping an event transversely in one or both views produces an equally valid event.  We use such flipping to effectively quadruple the size of the library when the matching is performed.  Each library event is used in each of the four possible configurations, and the best of the four is retained.  This symmetry is not quite perfect in the \nova{} detectors.  Attenuation in the readout fibers leads to subtly different charge resolutions and threshold effects on transversely opposing sides of an event, and NuMI neutrinos at the Far Detector enter at a $3^\circ$ upwards angle.  Nevertheless, the best-scoring matches come from the four possible flipped configurations with nearly equal probability: 26\% from unflipped events, 50\% from events with either one of the two views flipped, and 24\% from events with both views flipped.

\section{Decision tree}\label{sec:mva}

As library size increases, the fraction of an event's best matches that are truly signal tends toward the probability that the trial event itself is signal.  Further, all of the information available in the trial event is used when determining this probability.  It is in this sense that LEM is optimal.

For a library of finite and practical size, though, this signal fraction alone does not contain the full information extractable.  Other statistics constructed from the details of the best matches may, for example, indicate that the matches are drawn from an area of sparse library coverage and are thus less reliable.  The most powerful approach given a finite library is to construct several statistics describing the matches and to feed these into one of the standard multivariate analysis techniques to extract the final classifier.  In LEM, five variables are constructed from the 1,000 best library matches and are used as inputs to a decision tree, along with the calorimetric energy of the trial event as a sixth input.

\subsection{Weighted fraction of signal matches}

The basic quantity measuring what fraction of the best matches are signal events can be improved upon by weighting up the truly best matches over the lesser ones when calculating the signal fraction.  We use the weighting
\begin{equation}
  w^\prime_n = \exp\left(-\lambda\left({E_n \over E_{1000}}\right)^\gamma\right) \label{eqn:weightprime}\ ,
\end{equation}
where $n$ is the match index, $E_n$ is the energy of the $n^{\rm{th}}$ best match for the trial event, and $E_{1000}$ is the energy of the final (1000$^{\rm{th}}$) best match.  The optimized values used for $\lambda$ and $\gamma$ in \nova{} are
\begin{eqnarray}
  \lambda &=& 6.67\\
  \gamma &=& 10\ .
\end{eqnarray}
The typical ratio of weights $w^\prime_{1000}/w^\prime_{1}$ is $\mathord{\sim}0.1\%$, indicating that the most important matches are captured within the first thousand.

In practice, the weight must also include the oscillation probabilities alluded to earlier:
\begin{equation}\label{eqn:fullweight}
  w_n = w^\prime_nP^{\mathrm{osc}}_n\ ,
\end{equation}
where $P^{\mathrm{osc}}_n$ is the oscillation probability of match $n$, as described in \ref{sec:oscweights}.

All sums below that are indexed by $n$ run over the match list.  For notational convenience we also define $W\equiv\sum_n w_n$.  This weighting scheme is used for all five quantities formed from the best-match list.  The first is the weighted fraction of signal matches,
\begin{equation}
f_{\mathrm{sig}} = \frac{1}{W}\sum_{n,\;\mathrm{sig}}w_n\ ,
\end{equation}
where this sum includes only those terms due to signal matches.

\subsection{Mean hadronic $y$}

Signal events in which the outgoing electron carries only a small fraction of the incident neutrino's energy will look very much like NC background events.  The kinematic quantity $y$ (or rather, $1-y$) measures this fraction: $1-y=K_e/K_\nu$, where we've used $K_e$ and $K_\nu$ as the outgoing and incoming lepton energies to avoid confusion with the match energies $E$.   If a trial event matches well to signal events with high $y$, this can suggest that the trial event is in fact a high-$y$ NC event.  A second input is the mean $y$ for the best matches:
\begin{equation}
  \langle y\rangle = \frac{1}{W}\sum_nw_ny_n\ .
\end{equation}

\begin{figure*}
  \ifdefined\BW
    \includegraphics[width=.33\linewidth]{caf_pidexp_bw}
    \includegraphics[width=.33\linewidth]{caf_meanyexp_bw}
    \includegraphics[width=.33\linewidth]{caf_qfracexp_bw}\\
    \includegraphics[width=.33\linewidth]{caf_ediff_bw}
    \includegraphics[width=.33\linewidth]{caf_enrichexp_bw}
    \includegraphics[width=.33\linewidth]{caf_calE_bw}
  \else
    \includegraphics[width=.33\linewidth]{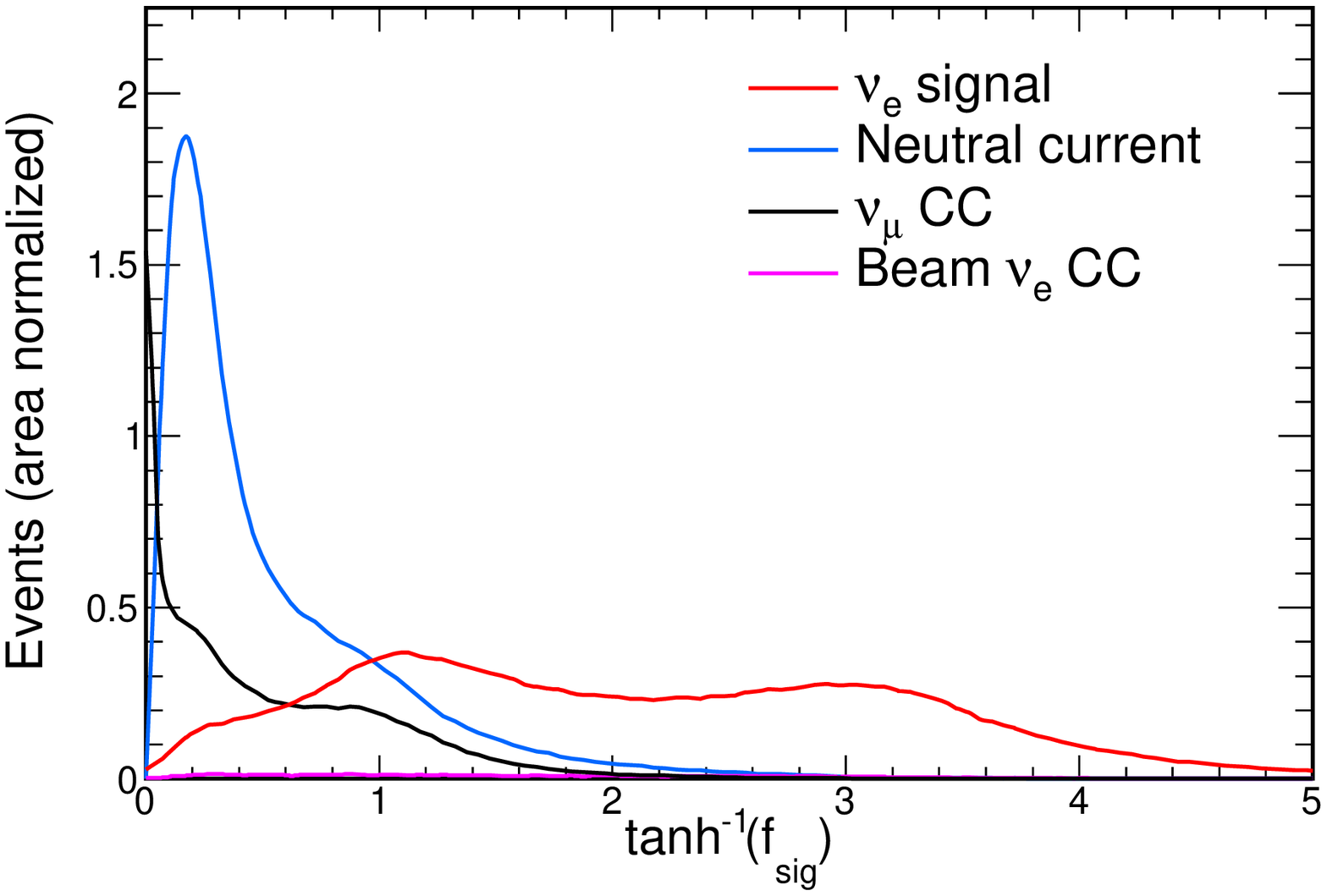}
    \includegraphics[width=.33\linewidth]{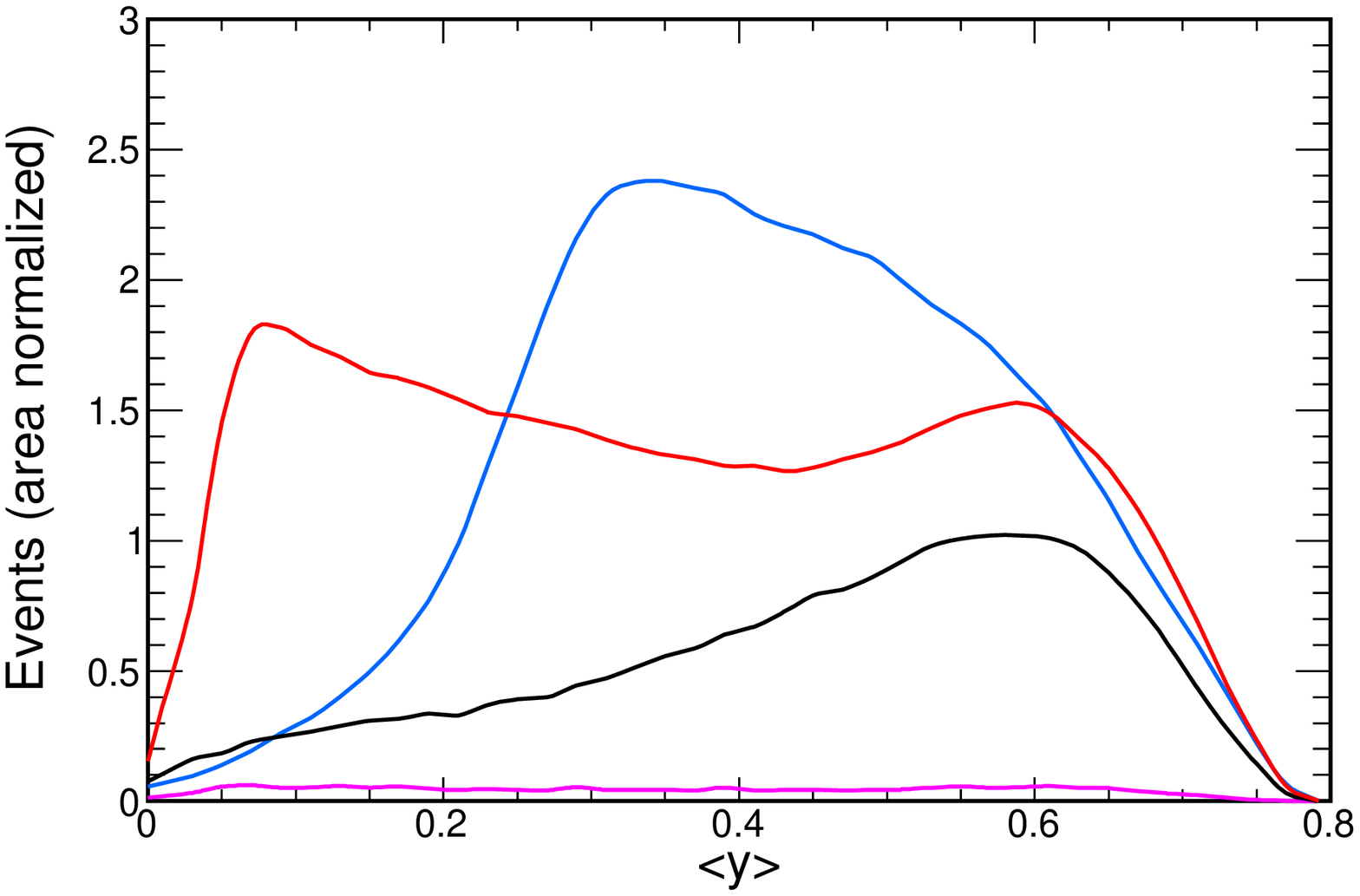}
    \includegraphics[width=.33\linewidth]{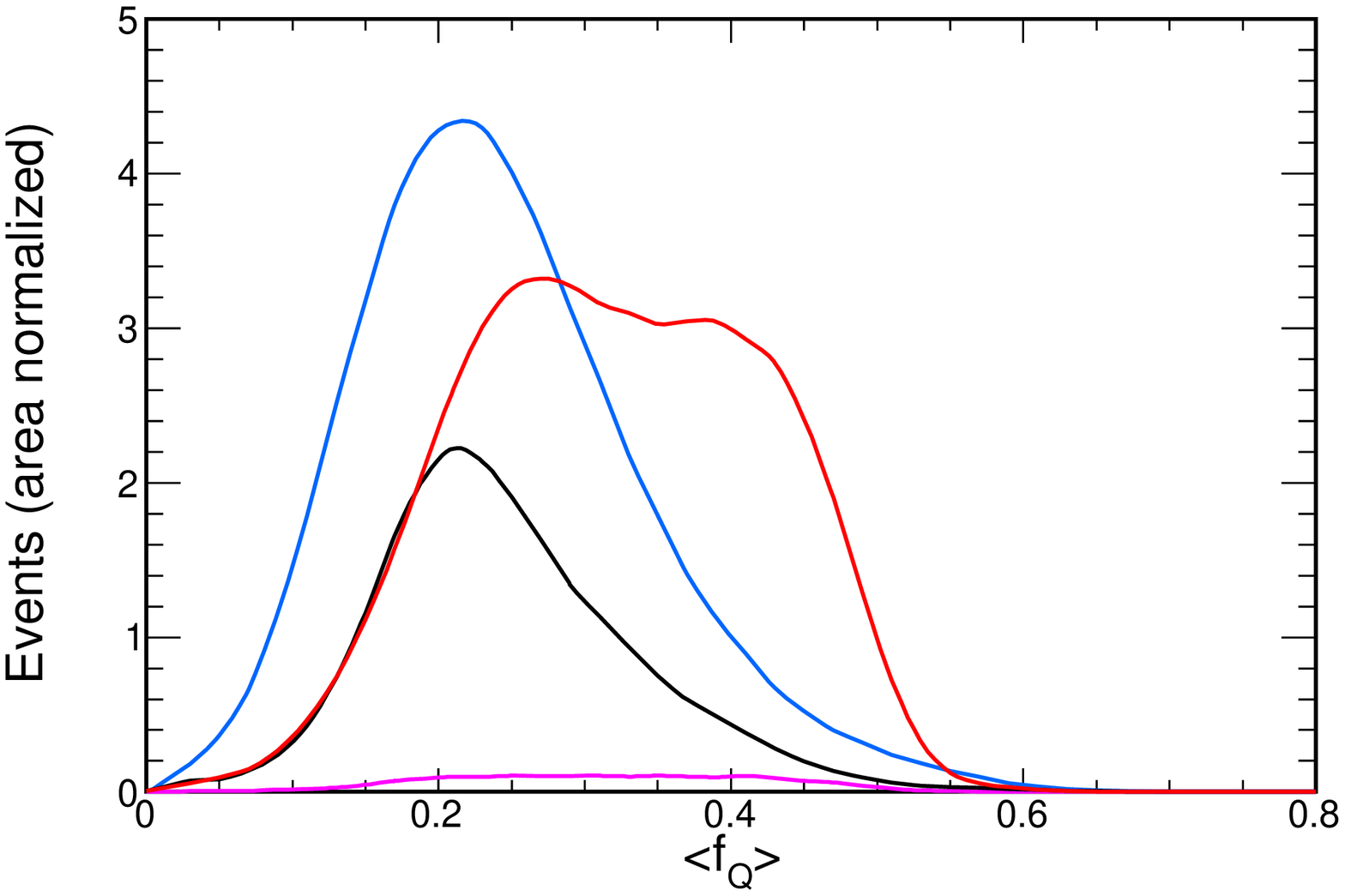}\\
    \includegraphics[width=.33\linewidth]{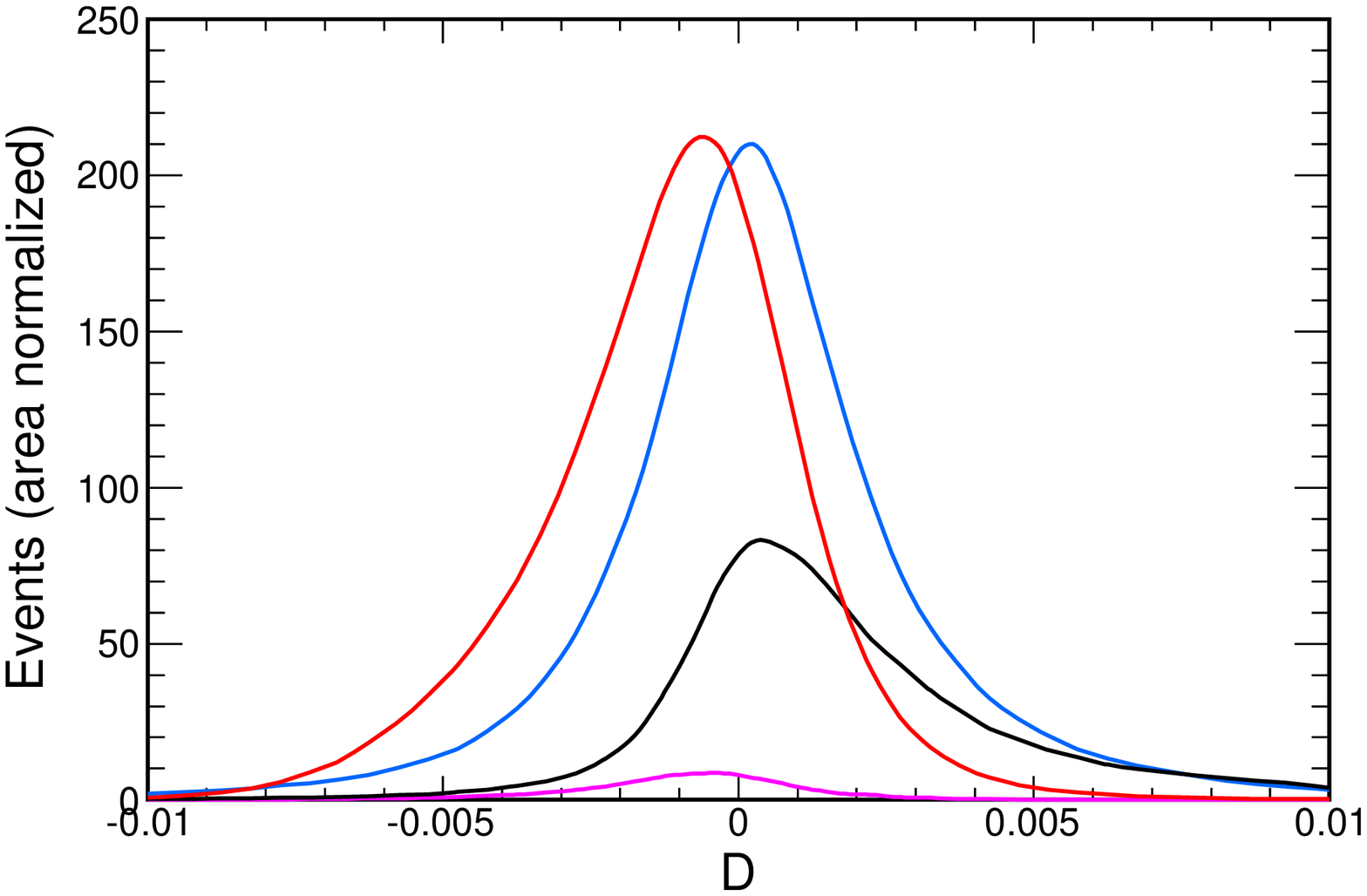}
    \includegraphics[width=.33\linewidth]{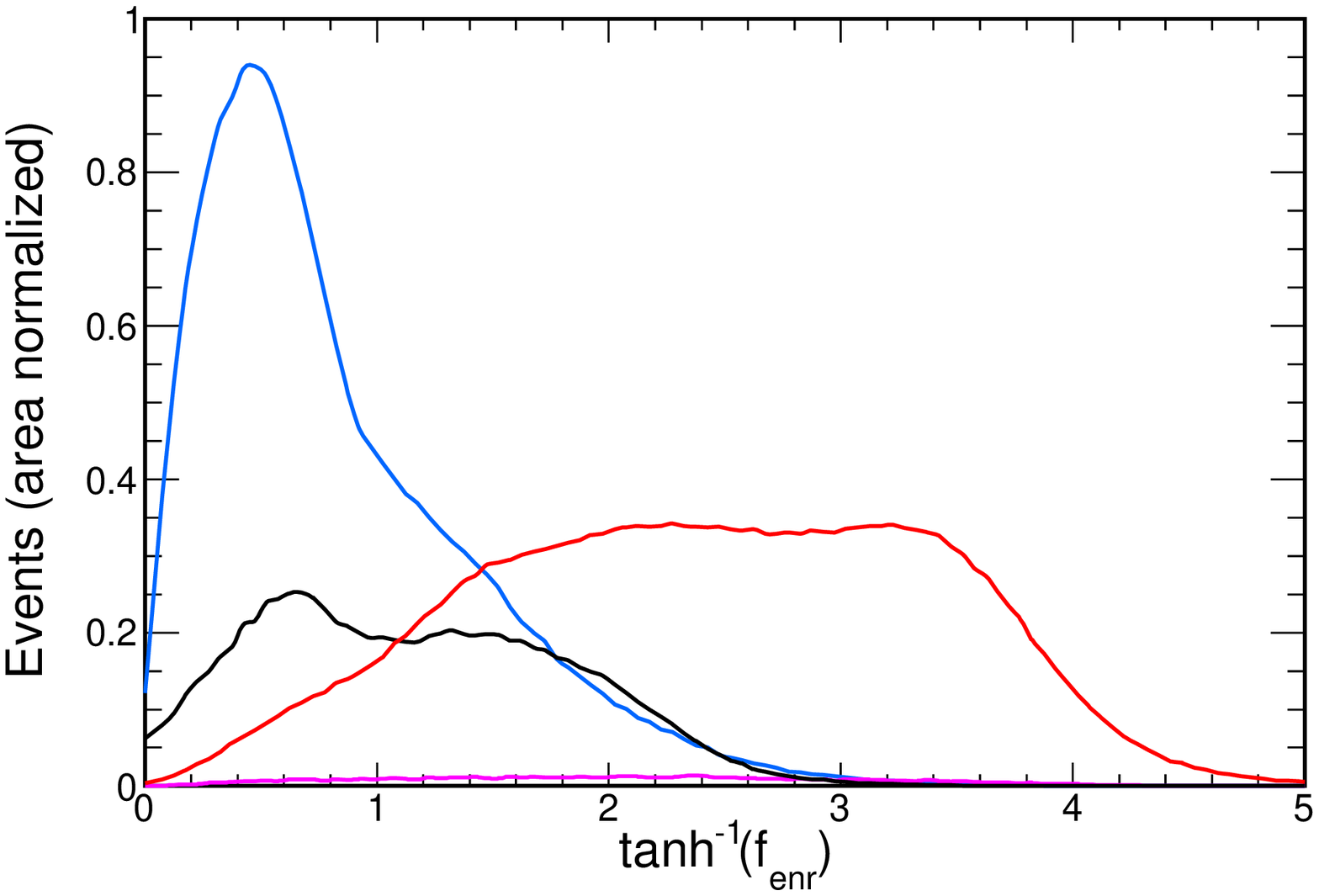}
    \includegraphics[width=.33\linewidth]{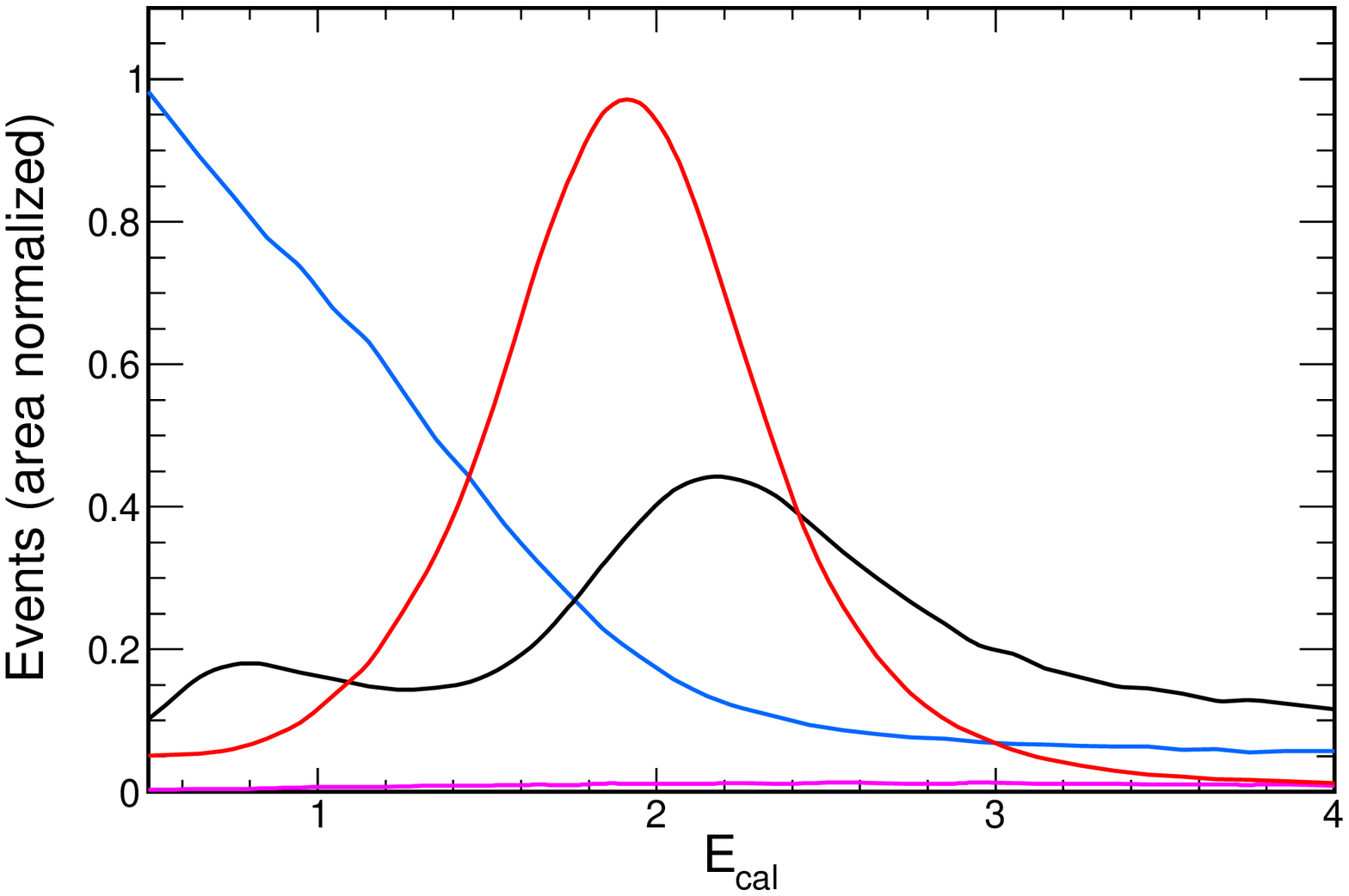}
  \fi
  \caption{The six decision tree inputs described in the text. The
    \sigstyle{} curves show the distribution of signal events. The \ncstyle{}, \numustyle{}, and
    \beamstyle{} curves show the distributions of neutral current,
    $\nu_\mu$ CC, and intrinsic $\nu_e$ CC backgrounds respectively. The $\nu_e$ signal
    and neutral current background are normalized to equal area. The other
    backgrounds are to the same scale as the neutral current
    curve.  The signal distributions for $f_{\mathrm{sig}}$ and $f_{\mathrm{enr}}$ are very sharply peaked at 1, so we have plotted these quantities as $\tanh^{-1}(f_{\mathrm{sig}})$ and $\tanh^{-1}(f_{\mathrm{enr}})$ to keep the signal and background curves visible on the same vertical scale.}\label{fig:vars}
\end{figure*}

\subsection{Mean matched charge fraction}

Matched charge fraction is an independent measure of the quality of the library matches, separate from the match energy.  For each trial/match pair, this is the quantity of charge that has a counterpart on identical cells in the two events divided by the total charge in the two events:
\begin{equation}
  f_Q = \frac{2\sum_i^{\rm cells} \min(a_i, b_i)}{\sum_i^{\rm cells}(a_i + b_i)}\ .
\end{equation}
The weighted average of the matched charge fraction over all the matches yields the next input:
\begin{equation}
  \langle f_Q\rangle = \frac{1}{W}\sum_nw_nf_{Q,n}\ .
\end{equation}

\subsection{Match energy difference}

This quantity measures whether the signal or background matches are the better matches on average.  It is the difference of the weighted mean energy of each class of matches:
\begin{equation}
  D \ =\  \frac{\sum_{n,\;\mathrm{sig}}w_nE_n}{\sum_{n,\;\mathrm{sig}}w_n} - \frac{\sum_{n,\;\mathrm{bkg}}w_nE_n}{\sum_{n,\;\mathrm{bkg}}w_n}
\end{equation}

\subsection{Enriched fraction}

The final match list quantity, similar in construction to $f_{\mathrm{sig}}$, is the weighted fraction of signal matches present among the signal and $\pi^0$-enriched matches ({\em i.e.}, excluding the non-enriched background)\ ,
\begin{equation}
  f_{\mathrm{enr}} = \frac{\sum_{n,\;\mathrm{sig}} w_n}{\sum_{n,\;\mathrm{enr}} w_n + \sum_{n,\;\mathrm{sig}} w_n}\ .
\end{equation}

\subsection{Total calorimetric energy}

NC backgrounds skew heavily to low visible energy thanks to the energy removed by the exiting neutrino.  The sum of all depositions $\{a_i\}$ recorded in the trial event,
\begin{equation}
  E_{\rm cal} = \sum_i^{\rm cells}a_i\ ,
\end{equation}
is included as a final input so that the classifier knows the prior expectations of signal and background.

\subsection{Choice of a decision tree, and figure of merit}

There are many multivariate techniques capable of combining these six input quantities into a single classifier output.  We investigated artificial neural networks, support vector machines, and decision trees.  An ensemble decision tree yielded the best performance of the approaches tried.  One problem with other techniques is that the figure of merit (f.o.m.)\ that, for example, artificial neural network training aims to minimize is the mean-squared-error of the classifier variable $c$:
\begin{equation}
  {\rm f.o.m.} \ =\  \sum_i^{\rm sig}(1-c)^2+\sum_i^{\rm bkg}c^2\ ,
\end{equation}
where the sums run over the signal and background training samples.  However, the figure of merit relevant to an experiment measuring the magnitude of a signal excess $s$ over a background $b$ with Poisson fluctuations is\begin{equation}
  {\rm f.o.m.}\ =\ {s\over\sqrt{s+b}}\ .
  \label{eqn:fom}
\end{equation}
If events are binned according to, say, the classifier output, the generalization is simply to sum in quadrature the
significances in the individual bins:
\begin{equation}
  {\rm f.o.m.}\ =\ \sqrt{\sum_i^{\rm bins}{s_i^2\over s_i+b_i}}\ .
  \label{eqn:binned_fom}
\end{equation}
While training a decision tree classifier, if the sample is divided at each step into subsamples 1 and 2 so as to maximize
\begin{equation}
  {s_1^2\over s_1+b_1}+{s_2^2\over s_2+b_2} ,
\end{equation}
then the performance of the full classifier is trivially optimized with respect to the figure of merit in Eq.~(\ref{eqn:binned_fom}).

The final classifier output is a voting ensemble of 1,000 decision trees each trained on a randomly chosen half of the full training sample.  The ensemble technique protects against over-training, a feature that we confirmed by evaluating the classifier performance on independent control samples.

\section{Classification performance}
\begin{figure}
  \ifdefined\BW
    \includegraphics[viewport=0 0 525 353,clip=true,width=\linewidth]{lem_spectrum_bw}
  \else
    \includegraphics[viewport=0 0 525 353,clip=true,width=\linewidth]{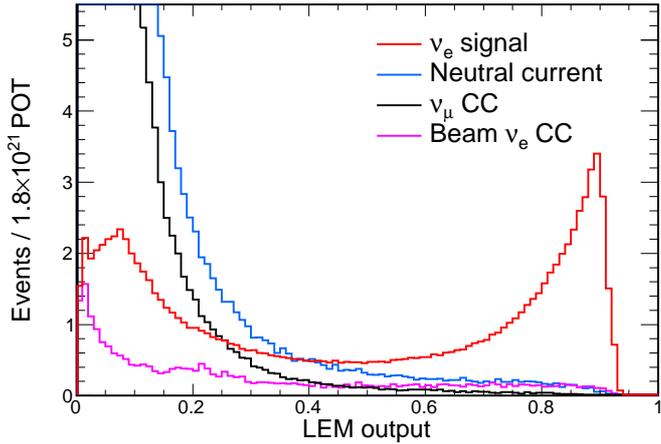}
  \fi

  \caption{The distribution of the LEM output variable for $\nu_e$ CC signal
    events (\sigstyle{}) compared to the background components:\ neutral current (\ncstyle{}),
    $\nu_\mu$ CC (\numustyle{}) and intrinsic beam $\nu_e$ CC (\beamstyle{}).  In
    order to make the details in the signal-like region visible, the $y$-axis truncates much of the
    background peak. 95\% of neutral current events and 98\% of
    $\nu_\mu$ charged current events have $\mathrm{LEM}\,\mathord{<}\,0.15$.  The distributions are scaled to a nominal 3-year NuMI exposure~\cite{nova} of $1.8\mathord{\times}10^{20}$~protons-on-target.}
  \label{fig:spectrum}
\end{figure}

\begin{figure}
  \includegraphics[viewport=0 0 525 358,clip=true,width=\linewidth]{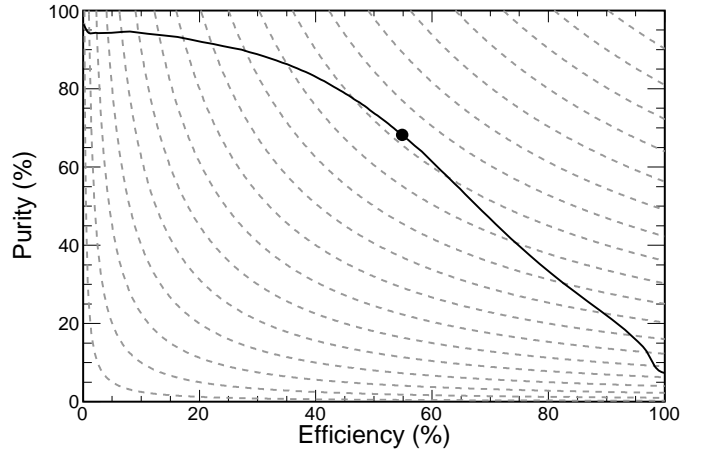}
  \caption{Efficiency and purity of the $\nu_e$ candidate sample selected by LEM for different cut positions. The dashed lines are curves of constant f.o.m.~$=s/\sqrt{s+b}$, and the solid circle indicates the result of the optimum cut.}
  \label{fig:eff_pur}
\end{figure}

Figure~\ref{fig:vars} shows the distribution of the six input variables for all event classes in the \nova{} $\nu_\mu\,\mathord{\to}\,\nu_e$ analysis.  Figure~\ref{fig:spectrum} shows the final LEM classifier output.  Figure~\ref{fig:eff_pur} shows the signal efficiency and purity obtained with various cuts on the LEM output.  All curves come from Monte Carlo simulation of the expected \nova{} data set.  We choose the cut on the LEM output variable that maximizes the figure-of-merit in Eq.~(\ref{eqn:fom}).  When applying LEM in a full experimental setting, one can fit the output distribution to gain additional discrimination power.

Table~\ref{tbl:counts} shows the expected number of signal and background events selected by the optimum LEM cut.  The signal efficiency is 55\% for a background mis-identification rate of 2.0\%.  The muon track of $\nu_\mu$ CC events keeps their mis-identification rate particulary low.  Background beam $\nu_e$ events are selected with a lower efficiency than signal $\nu_e$ events.  This is possible due to the different underlying energy spectra of the two classes.  As there is no absolute metric by which to judge the performance of the LEM classification algorithm described here, we note simply that the performance shown is excellent for the physics goals of \nova{}~\cite{nova}.

\begin{table*}
  \begin{centering}
    \begin{tabular}{lcc|cccc}
      & $\nu_e$ signal & Tot. bkgd. & NC & $\nu_\mu$ CC & Beam $\nu_e$ CC\\
      \hline
      No selection & 105     & 1332     &  734     & 573     & 25   \\
      LEM          &  58     &   27     &   14     &   4.6   &  7.9 \\
      \hline
      Efficiency   &  55\%   &    2.0\% &    2.0\% &    0.8\% & 32\%
    \end{tabular}\\
  \end{centering}
  \caption{Number of events expected in each event category initially and again after an optimal LEM cut assuming a nominal 3-year NuMI exposure of $1.8\times10^{21}$ protons-on-target.  The background is shown both as a total as well as broken down into NC, $\nu_\mu$ CC, and intrinsic beam $\nu_e$ CC components.  The bottom row shows the efficiencies for selecting events in each category.  The ``no selection'' row and the efficiencies derived from it count only those events with reconstructed visible energy between 0.5~GeV and 4~GeV.}
  \label{tbl:counts}
\end{table*}

\section{Computational optimization}\label{sec:comp}

\subsection{Speed}\label{sec:heads}

While each individual energy calculation can be performed very quickly, classifying a single event takes some time given the large size of the library.  For the \nova{} application, a single event must be treated in a second or so, which is the time scale required by other steps already performed during \nova{} event processing.  Without specialized hardware to run the inner loop, techniques to manage the LEM matching time focus on reducing the number of energies that need to be calculated.

We achieve a significant speed-up by introducing a library ``index''.  If trial event $A$ matches well to library event $B$, $A$ will likely match well to other library events that are, themselves, good matches to $B$.  Similarly, if $A$ and $B$ match poorly, then $A$ will likely match poorly to library events similar to $B$.

A library index is formed by drawing 10,000 events uniformly from the full library and matching each of these to the full library.  For each index event, a list of its 1,000,000 best-matched library events is saved.  This process happens ahead of time, at library creation.  When a trial event is classified, it is compared first to the 10,000 index events to find the single best-matching index event.  The trial event is then compared only to the 1M sibling events of that index event, reducing the total number of energies calculated per trial event from 77,000,000 to 1,010,000 -- a significant speed improvement that takes the per trial matching time from 97~s down to 1.7~s on a 2.3~GHz AMD Opteron processor.  Empirically, we find that 85\% of the trial event's ``true'' one-thousand top matches are captured with this indexed approach, and we find no noticeable degradation in the physics performance.

\subsection{Memory}

The speed optimization above is what allows the use of a 77M event library.  However, such a large library strains memory resources.  The full library is too large ($\sim$53~GB each for the library and index) to read from disk for each event, yet it is larger than the typical per-core memory allocation on grid computing nodes.

Thus, the library is converted from its original high-level format into the memory representation used by a running job.  This representation includes the self-energy of each event.  The conversion inflates the library slightly to 131~GB, but the advantage is that it can now be shared between running processes.  Each parallel matching job uses the {\tt mmap()} system call to make the contents of this file visible in its address space.  The mapping is marked read-only, so the kernel shares the pages between all the running processes. For example, on a 64-core server, the memory requirement to run 64 matching jobs is still only 131~GB, equivalent to an unshared 2~GB per core.  In case of memory pressure, the kernel will discard pages, knowing that they can be retrieved from disk (that is, the library file essentially acts as swap space) although this will significantly impact performance.

\section{Other information available in the match list}

In addition to signal-or-background classification, the detailed truth information available in the list of best matches
allows other information about the trial event to be inferred. One
could extract probabilities for different interaction modes, the inelasticity,
and so on, without requiring any independent reconstruction. An application
that has been pursued is the estimation of the incident neutrino energy for $\nu_e$ CC events.  Simply by averaging the true
neutrino energies of the best signal library matches and calibrating the resulting estimator, we achieve an energy resolution of 8.8\% on signal events selected by the oscillation analysis, competitive with other energy estimators in \nova{}.

\section{Summary}

The Library Event Matching algorithm compares input trial events to a large library of known events using all the information available, making LEM an optimal classifier given a sufficiently large library.  The \nova{} implementation of LEM has demonstrated excellent performance in separating $\nu_e$ signal from the key backgrounds, and a few simple optimizations have maintained practical computational requirements despite the large number of library events used.  Within the \nova{} context, the LEM technique has potential applications from reconstruction of the hadronic system to the event energy measure described above.  More broadly, LEM can be applied to completely different particle detectors or imaging systems in an array of fields and industries, wherever one needs to classify fine-grained images of objects whose visual characteristics vary in known ways.

\section{Acknowledgments}
The authors thank the \nova{} collaboration for use of its Monte Carlo simulation software and related tools.  This work was supported by the the US DOE under award DE-SC0006543.

\appendix

\section{Additional technical notes}
A few technical notes are included in this Appendix so as not to break up the discussion in the main text.

\subsection{Relation to other classification techniques}\label{sec:knnsvm}

If $f_{\rm sig}$ and $f_{\rm enr}$ were calculated unweighted, then those
variables would be $k$-nearest-neighbors classifiers, albeit with very large input
vectors.  With the weights $w_n$ applied, they act as kernel density
estimators. Note that
\begin{equation}
  \sqrt{2E} = \sqrt{\sum_{ij}\left(a_i^\beta-b_i^\beta\right)T_{ij}\left(a_j^\beta-b_j^\beta\right)}
\end{equation}
is a metric for the space of possible event images. That is, distances defined in
this way obey the triangle inequality.  For a Gaussian kernel in this space one
would expect $w_n\,\mathord{\sim}\,\exp(-E)$, which contrasts with the optimal value
of $\gamma\,\mathord{=}\,10$ found in practice.
Similarly $\langle y\rangle$ is an estimator for the true value of $y$ using the same kernel.

Methods exist to efficiently find nearest-neighbors in general metric
spaces without having to rely on heuristics such as the library index
in Section~\ref{sec:heads}.  Testing of a vantage-point
tree~\cite{vptree} indicated its performance was affected by the curse
of dimensionality.  A large fraction of the nodes would have to be
entered during a typical search.

\subsection{Energy calculation when $r_{ij}=0$}\label{sec:zeror}

The transfer matrix element $T_{ij}$ as written in Eq.~(\ref{eqn:transfer}) diverges when $i\,\mathord{=}\,j$ since $\Delta p_{ii}$ and $\Delta c_{ii}$ are zero.  Thus, for nearby cell pairs ($\Delta p_{ij}\,\mathord{\leq}\,5$ and $\Delta c_{ij}\,\mathord{\leq}\,5$), the energy calculation is performed as if the charge is distributed uniformly over each cell, with
\begin{equation}
  T_{ij} = \int\limits_0^1\!\!\!\int\limits_0^1\!\!\!\int\limits_0^1\!\!\!\int\limits_0^1
    \left[r_{ij}(x,y,u,v)\right]^{-\alpha}
       dx\,dy\,du\,dv\ ,
\end{equation}
where $(x, y)$ and $(u, v)$ scan over the areas of cells $i$ and $j$ and where $r_{ij}$ here is a generalization of the discrete distance used in the main text:
\begin{equation}
r_{ij}(x,y,u,v)=\sqrt{\left({\Delta p_{ij}+x-u\over\sigma_p}\right)^2+\left({\Delta c_{ij}+y-v\over\sigma_c}\right)^2}\ .
\end{equation}
For more distant pairs the simplified form of the transfer matrix given in Eq.~(\ref{eqn:transfer}) is sufficient.

\subsection{Neutrino oscillation weights}\label{sec:oscweights}
The retained matches are weighted according to Eq.~(\ref{eqn:fullweight}), which includes the probability for flavor oscillation.  The probabilities used are
\begin{eqnarray}
  P(\nu_\mu\,\mathord{\rightarrow}\,\nu_e)   &=& \sin^2\theta_{23}\sin^22\theta_{13}\sin^2\left({1.27\Delta m^2L\over E}\right)\\
  P(\nu_e\,\mathord{\rightarrow}\,\nu_\mu) &=& \sin^2\theta_{23}\sin^22\theta_{13}\sin^2\left({1.27\Delta m^2L\over E}\right)\\
  P(\nu_\mu\,\mathord{\rightarrow}\,\nu_\mu) &=& 1-\sin^22\theta_{23}\sin^2\left({1.27\Delta m^2L\over E}\right)\\
  P(\nu_e\,\mathord{\rightarrow}\,\nu_e) &=& 0\ ,
\end{eqnarray}
where $L\,\mathord{=}\,810~\mathrm{km}$ is the oscillation baseline, $E$ is the neutrino energy in GeV, and the oscillation parameters are taken to be
\begin{eqnarray}
\theta_{13}&=&9.2^\circ\\
\theta_{23}&=&38.5^\circ\\
\Delta m^2&=&2.35\times10^{-3}\ \mathrm{eV}^2\ .
\end{eqnarray}
These oscillation probabilities are first-order approximations to the full expressions.  This is both for practical reasons -- the second-order effects are poorly determined and are in fact what \nova{} aims to measure -- and because there is no requirement for the library have any particular distribution of events in it.  The second order effects can pull the probabilities higher or lower, making this weighting a reasonable middle ground for the library.  The library is also made devoid of intrinsic $\nu_e$ from the NuMI beam by setting that survival probability to zero.  The overall prefactor on the $\nu_\mu\,\mathord{\rightarrow}\,\nu_e$ (signal) line relative to the background lines actually does not enter in practice since the signal, background, and $\pi^0$-enriched background classes are scaled to have equal total weight in the library.

\end{document}